\pdfoutput=1
\documentclass[12pt]{article}
\usepackage[utf8]{inputenc}
\usepackage[british]{babel}
\usepackage{cmap}
\usepackage{lmodern}

\usepackage{amssymb, amsmath, amsthm}
\usepackage[a4paper,top=25mm,bottom=25mm,left=25mm,right=25mm]{geometry}
\usepackage{ragged2e}

\usepackage{authblk} 
\usepackage{pifont}
\usepackage{graphicx}
\usepackage[dvipsnames,svgnames,table]{xcolor}
\usepackage[figuresright]{rotating}
\usepackage{xtab} 
\usepackage{longtable} 
\usepackage{multirow}
\usepackage{footnote}
\usepackage[stable]{footmisc}
\usepackage{chngpage} 
\usepackage{pdflscape} 
\usepackage{tocbibind} 

\usepackage{pgfplots}
\pgfplotsset{compat=1.18}
\pgfplotsset{every tick label/.append style={font=\footnotesize}}
\usepackage{setspace}

\makesavenoteenv{tabular}
\usepackage{tabularx}
\usepackage{booktabs}
\usepackage{threeparttable} 
\usepackage[referable]{threeparttablex} 
\newcolumntype{R}{>{\raggedleft\arraybackslash}X}
\newcolumntype{L}{>{\raggedright\arraybackslash}X}
\newcolumntype{C}{>{\centering\arraybackslash}X}
\newcolumntype{A}{>{\columncolor{gray!25}}C}
\newcolumntype{a}{>{\columncolor{gray!25}}c}

\newlength{\tablen}

\usepackage{dcolumn} 
\newcolumntype{.}{D{.}{.}{-1}}

\usepackage{tikz}
\usetikzlibrary{arrows, calc, matrix, patterns, positioning, trees}
\usepackage[semicolon]{natbib}
\usepackage[hyphens]{url}
\usepackage{hyperref} 
\hypersetup{
  colorlinks   = true,    		
  urlcolor     = blue,    		
  linkcolor    = blue,			
  citecolor    = ForestGreen,		  	
}
\usepackage{microtype}
\usepackage[justification=centering]{caption} 

\usepackage[labelformat=simple]{subcaption}

\DeclareCaptionLabelFormat{parenthesis}{(#2)}
\makeatletter
\renewcommand\p@subfigure{\arabic{figure}.}
\makeatother

\DeclareCaptionLabelFormat{parenthesis}{(#2)}
\makeatletter
\renewcommand\p@subtable{\arabic{table}.}
\makeatother

\usepackage{alphalph}

\def\addlegendimage{\csname pgfplots@addlegendimage\endcsname}

\usepackage{enumitem}

\setlist[itemize]{leftmargin=2.5\parindent}
\setlist[enumerate]{leftmargin=2.5\parindent}

\theoremstyle{plain}

\theoremstyle{definition}

\theoremstyle{remark}

\def\keywords{\vspace{.5em} 
{\noindent \textit{Keywords}: }}

\def\AMS{\vspace{.5em} 
{\noindent \textbf{\emph{MSC} class}: }}

\def\JEL{\vspace{.5em} 
{\noindent \textbf{\emph{JEL} classification number}: }}

\title{Competitive balance in the UEFA Champions League group stage: Novel measures show no evidence of decline}
\author{
\href{https://sites.google.com/view/laszlocsato}{L\'aszl\'o Csat\'o}\thanks{~Corresponding author \newline
E-mail: \emph{laszlo.csato@sztaki.hun-ren.hu} \newline
Institute for Computer Science and Control (SZTAKI), Hungarian Research Network (HUN-REN), Laboratory on Engineering and Management Intelligence, Research Group of Operations Research and Decision Systems, Budapest, Hungary \newline 
Corvinus University of Budapest (BCE), Institute of Operations and Decision Sciences, Department of Operations Research and Actuarial Sciences, Budapest, Hungary}
$\qquad \qquad$
\href{https://sites.google.com/view/doragretapetroczy}{D\'ora Gr\'eta Petr\'oczy}\thanks{~E-mail: \emph{apetroczy@metropolitan.hu} \newline
MNB Institute, Budapest Metropolitan University, Budapest, Hungary}
}
\date{\today}

\def\Dedication{
{\noindent
``\emph{Different measurements are of different use, and all lines of research into competitive balance have, to date, proven quite instructive. To ignore this is to forgo important insights into the behavior of competitive balance.}''\footnote{~Source: \citet[Abstract]{FortMaxcy2003}.}
}}

\begin{document}

\newgeometry{top=25mm,bottom=25mm,left=25mm,right=25mm}
\maketitle
\thispagestyle{empty}
\Dedication

\begin{abstract}
\noindent
Competitive balance, which refers to the level of control teams have over a sports competition, is a crucial indicator for tournament organisers. According to previous studies, competitive balance has significantly declined in the UEFA Champions League group stage over the recent decades. Our paper introduces alternative indices to investigate this issue. Two ex ante measures are based on Elo ratings, and four dynamic concentration indicators compare the final group ranking to reasonable benchmarks. Using these indices, we find no evidence of any long-run trend in the competitive balance of the UEFA Champions League group stage between the 2003/04 and 2023/24 seasons.
\end{abstract}


\AMS{62P20, 91B82}

\JEL{L11, Z20, Z21}

\clearpage
\restoregeometry

\section{Introduction} \label{Sec1}

One of the most prestigious football tournaments, the UEFA Champions League, has been organised in the same format over 21 years, which has been fundamentally changed from the 2024/25 season. In particular, the traditional first stage, where four teams have played a home-away round-robin tournament in eight independent groups, has been replaced by a single incomplete round-robin league \citep{DevriesereCsatoGoossens2025}.
Commenting on this decision, the president of the Union of European Football Associations (UEFA), Aleksander {\v C}eferin, said \citep{UEFA2022a}:
``\emph{We are convinced that the format chosen strikes the right balance and that it will improve the competitive balance and generate solid revenues that can be distributed to clubs, leagues and into grassroots football across our continent while increasing the appeal and popularity of our club competitions}.''
Indeed, the new competition design can be successful in improving competitive balance as shown by \citet{Gyimesi2024}.


Consequently, this reform might have been inspired by problems perceived around the competitive balance of the UEFA Champions League. Recent results from academic literature support this assumption: a robust decline has been observed in both the ex ante \citep{Triguero-RuizAvila-Cano2023} and ex post \citep{RamchandaniPlumleyMondalMillarWilson2023, Triguero-RuizAvila-Cano2023} competitive balance of the UEFA Champions League group stage over the last decades.

Nonetheless, competitive balance is widely recognised to be a multi-dimensional concept \citep{Sanderson2002, KringstadGerrard2007, GerrardKringstad2022}, and there exist several measures for it \citep{DoriaNalebuff2021, GerrardKringstad2023, ManasisNtzoufrasReade2022, PawlowskiNalbantis2019}. Our paper introduces and explains two alternative approaches to quantify competitive balance in the UEFA Champions League group stage.

The first is a standard ex ante competitive balance measure, which evaluates the strengths of the participating teams based on past performance. While UEFA has an official indicator of team strength called UEFA club coefficient, used for seeding in the UEFA Champions League \citep{Csato2020a}, there is a more accurate measure of team strength called Football Club Elo Ratings \citep{Csato2024c}. Using Elo ratings instead of UEFA club coefficients has the additional advantage that our ex ante competitive balance measure will have a probabilistic interpretation.

The second indicator, called dynamic concentration, is based on the idea of \citet{Groot2008} to measure the ranking turnover between two seasons of a league by Kendall's $\tau$ coefficient. This method is adopted to compare an ex ante ranking to the ex post ranking in the groups within the same season. Hence, our measure essentially applies a standard interpretation of between-season competitive balance to one season. The motivation is that the rules of the UEFA Champions League group stage have not created powerful incentives to increase the number of points once the final rank of a team has become fixed \citep{CsatoMolontayPinter2024, Guyon2022a}.

Based on these indices, we find no evidence of any decline in the competitive balance of the UEFA Champions League group stage between the 2003/04 and 2023/24 seasons. Therefore, if UEFA has chosen the barely used incomplete round-robin design of the UEFA Champions League with its inherent challenges and risks \citep{CsatoDevriesereGoossensGyimesiLambersSpieksma2025} because of the worsening trend in competitive balance, the decision-makers might have been misled.

Naturally, the proposed indicators may have their own disadvantages and shortcomings, and we do not claim that they should be preferred to existing indices. However, it would be unreasonable to dismiss them for the sake of the standard measures of competitive balance as ``\emph{there are no easy formulas for competitive balance success}'' \citep[p.~120]{Zimbalist2002}.

We think it is the responsibility of the academic community to consider as many measures as possible, which will be done in the following structure.
Section~\ref{Sec2} provides a concise literature review on measuring competitive balance. Section~\ref{Sec3} summarises some possible limitations of previous studies.
Our measures of competitive balance are introduced in Section~\ref{Sec4} together with a description of the underlying data. Section~\ref{Sec5} provides the results and Section~\ref{Sec6} discusses them. Finally, Section~\ref{Sec7} offers concluding remarks.



\section{Related literature} \label{Sec2}

The economic analysis of team sports is often centred around the issue of competitive balance since the seminal articles of \citet{Rottenberg1956} and \citet{El-HodiriQuirk1971}. 
\citet{FortMaxcy2003} distinguish two approaches. Analysis of competitive balance focuses on what has happened to competitive balance over time or as a result of rule changes. The second line of literature on competitive balance studies its effect on the stakeholders, especially fans, and usually involves testing the longstanding uncertainty of outcome hypothesis. 

In recent decades, various indicators have been developed to quantify the competitive balance of sporting contests. 
The survey of \citet{PawlowskiNalbantis2019} classifies these measures according to the dimension of time. Short-term competitive balance refers to the uncertainty of individual games, often measured by using winning probabilities. The mid-term dimension focuses on seasonal uncertainty, while the long-term dimension concentrates on inter-seasonal uncertainty, the trend of competitive balance across several seasons. Our paper deals only with long-term competitive balance.

One commonly used measure of long-term competitive balance is the dispersion of winning percentage within sports leagues, also known as the Noll--Scully ratio \citep{Noll1988, Scully1989}. The Competitive Balance Ratio proposed by \citet{Humphreys2002} is an alternative measure that captures season-to-season changes in relative standings.

The Noll--Scully ratio is biased if leagues with different season lengths are compared. To address this issue, \citet{DoriaNalebuff2021} have introduced new indices that are invariant to season length, which has led to changes in competitive balance rankings among some major sports leagues, with surprising results such as the NFL moving from the most balanced to the least balanced league. The current study offers a similar re-evaluation of a ``common belief'' on competitive balance in the UEFA Champions League. 

The other popular concept to measure long-term competitive balance is the Herfindahl--Hirschman index ($\mathit{HHI}$) from the literature of industrial organization \citep{Herfindahl1950, Hirschman1945, Hirschman1964}, firstly used in sports by \citet{Depken1999}. The normalised version of $\mathit{HHI}$ is even more widespread \citep{OwenRyanWeatherston2007}. \citet{Triguero-RuizAvila-Cano2019} construct a novel index based on the concept of distance, with a range in the unit interval, which is the square root of the normalised $\mathit{HHI}$. It solves the problem of limited cardinality of most indices and makes the interpretation of the differences between the levels of competitive balance possible. \citet{Avila-CanoRuiz-SepulvedaTriguero-Ruiz2021} determine the maximum concentration as a function of the number of teams and the scoring system. Analogously, \citet{Triguero-RuizOwenAvila-Cano2023} compute the minimal $\mathit{HHI}$ in sports leagues without ties.

Alternative indicators of long-term competitive balance have also been suggested.
\citet[Chapter~1.4]{Groot2008} proposes Kendall rank correlation to obtain a dynamic measure of competitive balance.
\citet{ManasisAvgerinouNtzoufras2011} consider the normalised concentration ratio to evaluate the extent to which a league is dominated by a small subset of teams.
\citet{ManasisAvgerinouNtzoufrasReade2013} introduce an index to capture the degree of competition for winning the important prizes awarded in European football leagues such as the championship, the qualification for prestigious international club competitions, and avoiding relegation.
\citet{ManasisNtzoufras2014} examine the applicability of between-season competitive balance indices in the context of European football, and define a set of specially designed measures.
\citet{HoodJewell2022} illustrate the value of using betting data to simulate an ex-ante distribution of league-point outcomes in English football.
\citet{BasiniTsouliNtzoufrasFriel2023} develop a statistical network model to assess the balance between teams in a league.
\citet{GerrardKringstad2023} extend previous empirical studies on the four North American major sports leagues using six competitive balance measures. Similar to us, they conclude that the evaluation of competitive balance is both metric-dependent and time-dependent, which highlights the importance of considering a portfolio of indices rather than a single indicator.
\citet{CsatoPetroczy2024} adopt bibliometric indices for quantifying competitive balance in the UEFA Champions League knockout stage.
A recent empirical analysis of nontransitive patterns in professional football \citep{vanOurs2024} can provide a novel approach to measuring competitive balance, too.

Traditional measures of performance have recently been challenged in multi-stage cycling races \citep{Ausloos2024a, Ausloos2024b}. Our paper has a similar aim by suggesting novel competitive balance measures in the group stage of (football) tournaments.

\section{Potential limitations of previous studies} \label{Sec3}

Competitive balance has two interpretations: ex ante and ex post.
Ex ante competitive balance is experienced before the matches are played and is connected to suspense; the consumers hope to see an exciting game where winning and losing are far from being predetermined \citep{RichardsonNalbantisPawlowski2023}.
On the other hand, ex post competitive balance is related to surprise. According to \citet{ElyFrankelKamenica2015}, it is more surprising if the current events are greatly different from the previous events.

Compared to national leagues, measuring competitive balance in the UEFA Champions League has an additional challenge, that is, a team participating in a given year may not compete in the following year but replaced by a stronger team from the same domestic league \citep{Koning2009}. Therefore, the strength of the teams should be quantified independently of their identity.

\citet{Triguero-RuizAvila-Cano2023} start to measure ex ante competitive balance by calculating the shares of the clubs from UEFA club coefficients. They are used to compute $\mathit{HHI}$ for each group, which gives the distance to competitive balance ($\mathit{DCB}$) index \citep{Dubois2022, ScellesFrancoisDermit-Richard2022, Triguero-RuizAvila-Cano2019} as follows:
\[
\mathit{DCB} = \sqrt{\frac{\mathit{HHI} - \mathit{HHI}_{\min}}{\mathit{HHI}_{\max} - \mathit{HHI}_{\min}}}.
\]

UEFA club coefficient is the official measure of team strength in UEFA club competitions, the number of points gained by a team in UEFA club competitions over the last five years \citep{DagaevRudyak2019}. There is a lower bound based on the five-season association coefficient, which might be relevant for emerging clubs without meaningful experience at the international level \citep{Csato2022b}.
Primarily, the rankings derived from the UEFA club coefficients have determined seeding in the UEFA club competitions \citep{Csato2020a, CsatoIlyin2025}. 

However, UEFA club coefficients suffer from a major shortcoming: they do not reflect the results of the majority of matches played by the clubs since all games in the national leagues and cups are omitted. Home advantage does not count either. Therefore, UEFA club coefficients usually underestimate emerging clubs in the top leagues without a recent international presence.
A good example is the English club Aston Villa in the 2024/25 UEFA Champions League, which had the 31st highest UEFA club coefficient but was ranked 20th according to its Elo rating on 2 September 2024.
In contrast, the Ukrainian champion Shakhtar Donetsk was 17th according to its UEFA club coefficient and 33rd according to its Elo rating.
Unsurprisingly, a variant of the Elo method provides much higher accuracy in terms of explanatory power for the UEFA Champions League \citep{Csato2024c}.

Ex post competitive balance is usually quantified by the Herfindahl--Hirschman index ($\mathit{HHI}$) based on the shares of each team from the total points awarded \citep{Avila-CanoOwenTriguero-Ruiz2023, OwenOwen2022, OwenRyanWeatherston2007}.
Both \citet{RamchandaniPlumleyMondalMillarWilson2023} and \citet{Triguero-RuizAvila-Cano2023} use this approach to compute ex post competitive balance in a group. The underlying assumption is that the teams have appropriate incentives to collect points even though their qualification depends on the group ranking rather than on the number of points. Thus, the clubs do not necessarily have appropriate incentives in the last round(s) \citep{ChaterArrondelGayantLaslier2021, CsatoMolontayPinter2024}.
\citet{BuraimoForrestMcHaleTena2022} find robust evidence that matches of no importance for seasonal outcomes---which is equivalent to group ranking in our setting---may not be important to viewers in the English Premier League.

In the history of football, numerous examples exist when a team was satisfied with a (moderate) loss if it was still sufficient to achieve its objectives \citep{Guyon2020a, Guyon2022a, KendallLenten2017}. This consideration could be especially important in the UEFA Champions League group stage since the clubs usually play in their domestic leagues both on the previous and subsequent weekends, which creates a powerful incentive to rest the best players and field a younger squad to gain experience if the game is stakeless with respect to the rank of the team.

As an illustration, consider Group G in the 2018/19 UEFA Champions League \citep{CsatoMolontayPinter2024}. Before the last matchday, Real Madrid already won the group, and Roma was guaranteed to finish in the second place. In the last round, Real Madrid suffered a shocking 3-0 defeat against CSKA Moscow after fielding a fully rotated squad, which was its first European home loss by more than two goals \citep{Bell2018}. Roma also lost against Viktoria Plze{\v n} despite its home win of 5-0 in the second round.
We have serious doubts that most consumers perceive the group to be more balanced---as implied by $\mathit{HHI}$---merely because Real Madrid lost rather than won its last stakeless match.


Nevertheless, in contrast to the distribution of the results obtained, the final group ranking cannot be evaluated in itself without any benchmark. In other words, the group ranking could not serve as an ex post competitive balance measure because it should be compared to an ex ante ranking.

Based on the arguments above, two indices of ex ante competitive balance and four indices of dynamic concentration will be constructed in the following. Naturally, our measures have their own weaknesses, and they should not necessarily be preferred to the indicators used in the previous literature.

\section{Methodology and data} \label{Sec4}

The UEFA Champions League was organised in the same format between the 2003/04 and 2023/24 seasons. The group stage contained eight groups of four teams each. The top two teams qualified for the Round of 16, where the group winners were matched with the runners-up, the third-placed teams were transferred to the second most prestigious UEFA club competition, and the fourth-placed teams were eliminated.

In order to ensure the balancedness of the groups, the 32 teams were allocated into four pots. Until the 2014/15 season, seeding was determined by the UEFA club coefficients except for assigning the titleholder to the first pot \citep{Csato2021a}. The first pot consisted of the titleholder and the champions of the highest-ranked associations from the 2015/16 season \citep{DagaevRudyak2019}, and the UEFA Champions League titleholder, the UEFA Europa League titleholder, as well as the champions of the highest-ranked associations from the 2018/19 season \citep{Csato2020a}.

Sections~\ref{Sec41} and \ref{Sec42} motivate and define the indices of ex ante competitive balance and dynamic concentration, respectively.
Section~\ref{Sec43} provides an example of their calculation, Section~\ref{Sec44} gives information about the underlying data, while Section~\ref{Sec45} presents how the dynamics of competitive balance is analysed.
Last but not least, Section~\ref{Sec46} summarises the main steps of our research.

\subsection{Measures of ex ante competitive balance} \label{Sec41}

Following the recent literature \citep{BoskerGurtler2024, Csato2024c, YildirimBilman2025a, YildirimBilman2025b}, we use Football Club Elo Ratings as a measure of team strength \citep{FootballClubEloRatings}. It gives the expected probability that team $i$ wins against team $j$ as:
\begin{equation} \label{Win_prob}
W_{ij} = \frac{1}{1 + 10^{- \left( R_i - R_j \right) / 400}},
\end{equation}
where $R_i$ and $R_j$ are the Elo ratings of the two teams, respectively. Obviously, $W_{ji} = 1- W_{ij}$.

To update the Elo ratings, the expected winning probability $W_{ij}$ is compared to the actual result $Q$ (1 for win, 0.5 for draw, 0 for loss):
\[
\Delta R_i = 20 \left( Q - W_{ij} \right).
\]
Since $\Delta R_i = - \Delta R_j$, the sum of Elo ratings does not change after any game.
Finally, home advantage is taken into account by increasing the Elo difference in \eqref{Win_prob}. The home advantage parameter is not fixed in advance but dynamically updated to guarantee that it converges to the actual effect of home advantage \citep{FootballClubEloRatings}.

Groups of four teams imply six pairs of clubs. For each pair, the winning probability of the stronger team---a value between 0.5 and 1---is computed according to \eqref{Win_prob}. The six values are added to get $\mathit{UCB}_1^A$, which is normalised similarly to the idea behind the $\mathit{DCB}$:
\[
\mathit{CB}_1^A = \frac{\mathit{UCB}_1^A - \left( \mathit{UCB}_1^A \right)_{\min}}{\left( \mathit{UCB}_1^A \right)_{\max} - \left( \mathit{UCB}_1^A \right)_{\min}} = \frac{\mathit{UCB}_1^A - 3}{6 - 3} = \frac{\mathit{UCB}_1^A}{3} - 1.
\]
$\mathit{CB}_1^A \in \left[ 0,1 \right]$ is our first index of ex ante competitive balance, for which a lower value is preferred since it indicates a higher (expected) uncertainty in the group matches.

If only two teams qualify from a group of four, a group is usually said to be a harsh ``group of death'' with three strong teams, independent of the strength of the fourth team \citep{Guyon2015a, LalienaLopez2019}. Thus, it makes sense to focus on the three strongest teams and derive the indicator from the three pairs implied. Our second index of ex ante competitive balance, normalised to lie between 0 and 1 is
\[
\mathit{CB}_2^A = \frac{\mathit{UCB}_2^A - \left( \mathit{UCB}_2^A \right)_{\min}}{\left( \mathit{UCB}_2^A \right)_{\max} - \left( \mathit{UCB}_2^A \right)_{\min}} = \frac{\mathit{UCB}_2^A - 1.5}{3 - 1.5} = \frac{2 \mathit{UCB}_2^A}{3} - 1.
\]
Naturally, $\mathit{CB}_1^A \geq \mathit{CB}_2^A$ always holds due to the definitions. However, we will only investigate the trends in the evolution of one particular measure, and never compare $\mathit{CB}_1^A$ to $\mathit{CB}_2^A$ directly.

It is worth noting that some papers have used similar indicators to quantify ex ante competitive balance.
\citet{HoodJewell2022} adopt betting odds to simulate an ex-ante distribution of league outcomes. \citet{Gyimesi2024} uses the numerical average of the absolute Elo difference of each match to measure the uncertainty of match outcomes in the new format of the UEFA Champions League.

To conclude, our approach for quantifying ex ante competitive balance has two advantages:
(1) it has a straightforward interpretation as the average (normalised) winning probability of the stronger team in each group match; and
(2) since Elo ratings imply a higher predictive power than UEFA club coefficients, consumers may show greater sensitivity to its fluctuation within the groups.
On the other hand, the number of teams per national association is allocated according to UEFA association coefficients, which are based on UEFA club coefficients. Furthermore, the competing teams are determined by the results of the different national leagues. Finally, the Elo rating is somewhat persistent and may not reflect the real strength of the team; however, this caveat also applies to the UEFA club coefficient.

Following \citet{Triguero-RuizAvila-Cano2023}, the ex ante measures of competitive balance $\mathit{CB}_1^A$ and $\mathit{CB}_2^A$ are averaged for the eight groups in each season.

\subsection{Measures of dynamic concentration} \label{Sec42}

We compute the Kendall rank correlation coefficient \citep{Kendall1938} between an ex ante ranking and the final group ranking.
The use of Kendall rank correlation for competitive balance goes back at least to \citet{Groot2008}, who applied it to evaluate dynamic competitive balance in football leagues. In a sense, competitive balance in the UEFA Champions League groups is about mobility: an ex ante (before the season) ranking of the teams can be compared to the final (after the season) group ranking to determine the uncertainty of tournament outcome. Mobility is maximal and dynamic concentration is minimal if the final ranking is exactly the reverse of the ex ante ranking, corresponding to a perfect negative rank correlation. However, then one can perfectly predict the final ranking by reversing the initial order of the teams. Hence, the ideal situation is if the final ranking is random, that is, Kendall rank correlation equals zero \citep{Groot2008}.

Denote the number of concordant and discordant pairs between the ex ante and ex post rankings by $P$ and $Q$, respectively, and the number of teams by $n$.
The Kendall $\tau$ rank correlation equals
\[
\tau = \frac{4P}{n(n-1)} - 1 = 1 - \frac{4Q}{n(n-1)}.
\]
The ex post ranking is, naturally, the final group ranking. However, two plausible choices exist for the ex ante rankings:
\begin{itemize}
\item
The initial pot allocation since each pot contains one team from each pot;
\item
The ranking implied by the Football Club Elo Ratings before the season.
\end{itemize}
In addition, the consumers may ignore if the group winner and the runner-up are swapped because both of them qualify for the Round of 16. Then the maximum of discordant pairs is $n(n-1)/2-1$, and the analogously modified rank correlation measure is
\[
\hat{\tau} = 1 - \frac{4 \hat{Q}}{n(n-1)-1},
\]
where $\hat{Q}$ is the adjusted number of discordant pairs such that a discordant pair in the first two positions is disregarded, while all other discordant pairs are retained.

\begin{table}[t!]
  \centering
  \caption{Indices of dynamic concentration}
  \label{Table1}
    \begin{tabularx}{\textwidth}{cCc} \toprule
    Measure & Ex ante ranking & Are group winners and runners-up distinguished? \\ \midrule
    $\mathit{DC}_1$ & Pot allocation & \textcolor{ForestGreen}{\ding{52}} (the formula of $\tau$ applies) \\
    $\mathit{DC}_2$ & Pot allocation & \textcolor{BrickRed}{\ding{55}} (the formula of $\hat{\tau}$ applies) \\
    $\mathit{DC}_3$ & Elo rating & \textcolor{ForestGreen}{\ding{52}} (the formula of $\tau$ applies) \\
    $\mathit{DC}_4$ & Elo rating & \textcolor{BrickRed}{\ding{55}} (the formula of $\hat{\tau}$ applies) \\ \bottomrule
    \end{tabularx}
\end{table}

The four alternative measures of dynamic concentration are shown in Table~\ref{Table1}.
Following \citet{RamchandaniPlumleyMondalMillarWilson2023} and \citet{Triguero-RuizAvila-Cano2023}, these measures of competitive balance are also averaged for the eight groups in each season.

\subsection{An illustrative example} \label{Sec43}

The four teams of Group C in the 2023/24 UEFA Champions League are Napoli (drawn from Pot 1; Elo rating: 1911), Real Madrid (Pot 2; 1917), Braga (Pot 3; 1677), and Union Berlin (Pot 4; 1757). The pairwise differences between their Elo ratings are 6, 234, 154, 240, 160, 80, respectively, thus
\begin{eqnarray*}
\mathit{UCB}_1^A & = & \frac{1}{1 + 10^{-6/400}} + \frac{1}{1 + 10^{-234/400}} + \frac{1}{1 + 10^{-154/400}} + \frac{1}{1 + 10^{-240/400}} + \\
& & + \frac{1}{1 + 10^{-160/400}} + \frac{1}{1 + 10^{-80/400}} = 4.138.
\end{eqnarray*}
Consequently, the stronger team wins a group match with a probability of 69\% in average.

Analogously, the differences between the three highest Elo ratings are 6, 154, 160, respectively, thus
\[
\mathit{UCB}_2^A = \frac{1}{1 + 10^{-6/400}} + \frac{1}{1 + 10^{-154/400}} + \frac{1}{1 + 10^{-160/400}} = 1.932.
\]
Consequently, the stronger team wins a group match with a probability of 64\% in average if we focus on the three favourites.

The corresponding normalised values are:
\[
\mathit{CB}_1^A = \frac{\mathit{UCB}_1^A}{3} - 1 = 0.379 \qquad \text{and} \qquad \mathit{CB}_2^A = \frac{\mathit{2UCB}_2^A}{3} - 1 = 0.288.
\]

The final group ranking is Real Madrid, Napoli, Braga, Union Berlin.
Compared to the pot allocation, one discordant pair (Napoli and Real Madrid) exists, hence $\mathit{DC}_1 = 1 - 4/12 = 2/3$.
However, this discordant pair affects the first two places in the group ranking, hence $\mathit{DC}_2 = 1 - 0/10 = 1$.
Compared to the ranking implied by Elo ratings, there is one discordant pair (Braga and Union Berlin), hence $\mathit{DC}_3 = 1 - 4/12 = 2/3$.
This discordant pair affects not only the first two places in the group ranking, hence $\mathit{DC}_4 = 1 - 4/10 = 3/5$.

\subsection{Data} \label{Sec44}

We consider all seasons between 2003/04 and 2023/24 when the tournament format of the UEFA Champions League did not change.
The pot allocation and group rankings are readily available from several sources; we have used Wikipedia after cross-checking with the official UEFA site.

Football Club Elo Ratings on a given date can be downloaded from \url{api.clubelo.com/YYYY-MM-DD}.
In contrast to UEFA club coefficients, the Elo rating of a team changes during the season. Thus, the Elo ratings on 1 September are used for each season because the group stage is usually played between September and December. The Elo ratings are rounded to the nearest integer. Then Group B in the 2020/21 UEFA Champions League contains two teams (Shakhtar Donetsk and Borussia M\"onchengladbach) with the same value (1766), which should be broken for indices $\mathit{DC}_3$ and $\mathit{DC}_4$. It is decided by the higher value before rounding, but the tie-breaking does not affect our qualitative results.

\subsection{The evolution of competitive balance} \label{Sec45}

The approach of \citet{RamchandaniPlumleyMondalMillarWilson2023} and \citet{Triguero-RuizAvila-Cano2023} is followed directly.
For all the six indicators of competitive balance, the linear trend is obtained by estimating a simple regression model:
\[
\mathit{I}_{j,t} = c_j + \alpha_j t + \varepsilon_t,
\]
where $\mathit{I}_{j,t}$ is the $j$th competitive balance measure and $\varepsilon_t$ is the error term in season $t$ ($2003 \leq t \leq 2023$). Every season is denoted by its first year when the group stage is played. Furthermore, $c_j$ is the intercept and $\alpha_j$ is the coefficient of the season.
Competitive balance declined/did not change/improved if $\alpha_j$ is significantly positive/statistically zero/significantly negative.

\subsection{Overview} \label{Sec46}

Our results can be replicated by the following procedure:
\begin{enumerate}
\item
Collecting the publicly available Football Club Elo Ratings, UEFA Champions League pot allocations and group rankings between the 2003/04 and 2023/24 seasons (see Section~\ref{Sec44} for details);
\item
Calculating competitive balance measures $\mathit{CB}_1^A$, $\mathit{CB}_2^A$ (see Section~\ref{Sec41} for details) and $\mathit{DC}_1$--$\mathit{DC}_4$ (see Section~\ref{Sec42} for details) for the eight groups in each season (see Section~\ref{Sec43} for details);
\item
Estimation of linear regressions to check the existence of any trend in the seasonal averages of $\mathit{CB}_1^A$, $\mathit{CB}_2^A$, and $\mathit{DC}_1$--$\mathit{DC}_4$ (see Section~\ref{Sec45} for details).
\end{enumerate}

\section{Results} \label{Sec5}

\begin{figure}[ht!]
\centering

\begin{tikzpicture}
\begin{axis}[
name = axis1,
title = {Ex ante measure $\mathit{CB}_1^A$},
title style = {font=\small},
xlabel = Season,
x label style = {font=\small},
x tick label style = {/pgf/number format/1000 sep={}},
ylabel = Value of competitive balance,
y label style = {font=\small},
width = 0.5\textwidth,
height = 0.4\textwidth,
ymajorgrids = true,
xmin = 2002.5,
xmax = 2023.5,
ymin = 0.21,
ymax = 0.51,
] 
\addplot [blue, mark=star, mark size=2pt, mark options=solid, only marks] coordinates {
(2003,0.283710492057469)
(2004,0.322395232984772)
(2005,0.417435545955688)
(2006,0.337471754696038)
(2007,0.419331606774874)
(2008,0.459135051038576)
(2009,0.434908310815382)
(2010,0.413231567489708)
(2011,0.426852549393698)
(2012,0.391918184090656)
(2013,0.411577365228621)
(2014,0.488245487093441)
(2015,0.461394017994427)
(2016,0.417641744994786)
(2017,0.461579405735787)
(2018,0.397559412091651)
(2019,0.38885569833192)
(2020,0.41598206351716)
(2021,0.406122790968469)
(2022,0.386253962137401)
(2023,0.351822524355051)
};
\draw [red,thick,dashed] (\pgfkeysvalueof{/pgfplots/xmin},-3.5418+0.00196*\pgfkeysvalueof{/pgfplots/xmin}) -- (\pgfkeysvalueof{/pgfplots/xmax},-3.5418+0.00196*\pgfkeysvalueof{/pgfplots/xmax});
\end{axis}

\begin{axis}[
at = {(axis1.south east)},
xshift = 0.1\textwidth,
title = {Ex ante measure $\mathit{CB}_2^A$},
title style = {font=\small},
xlabel = Season,
x label style = {font=\small},
x tick label style = {/pgf/number format/1000 sep={}},
ylabel = Value of competitive balance,
y label style = {font=\small},
width = 0.5\textwidth,
height = 0.4\textwidth,
ymajorgrids = true,
xmin = 2002.5,
xmax = 2023.5,
ymin = 0.21,
ymax = 0.51,
] 
\addplot [blue, mark=star, mark size=2pt, mark options=solid, only marks] coordinates {
(2003,0.252685084229457)
(2004,0.240113297385791)
(2005,0.349653783148035)
(2006,0.313759115600547)
(2007,0.354844732474096)
(2008,0.385735050071426)
(2009,0.30315093329707)
(2010,0.366295189246344)
(2011,0.321708291000443)
(2012,0.279763058286936)
(2013,0.375906626163598)
(2014,0.395420250491238)
(2015,0.333757293548977)
(2016,0.333106140352104)
(2017,0.357293498122064)
(2018,0.341036773697465)
(2019,0.333709015993894)
(2020,0.351170103827152)
(2021,0.310597443539946)
(2022,0.297383807009441)
(2023,0.327010697669181)
};
\draw [red,thick,dashed] (\pgfkeysvalueof{/pgfplots/xmin},-2.5841+0.001447*\pgfkeysvalueof{/pgfplots/xmin}) -- (\pgfkeysvalueof{/pgfplots/xmax},-2.5841+0.001447*\pgfkeysvalueof{/pgfplots/xmax});
\end{axis}
\end{tikzpicture}

\vspace{0.5cm}
\begin{tikzpicture}
\begin{axis}[
name = axis1,
title = {Dynamic concentration $\mathit{DC}_1$},
title style = {font=\small},
xlabel = Season,
x label style = {font=\small},
x tick label style = {/pgf/number format/1000 sep={}},
ylabel = Value of competitive balance,
y label style = {font=\small},
width = 0.5\textwidth,
height = 0.4\textwidth,
ymajorgrids = true,
xmin = 2002.5,
xmax = 2023.5,
ymin = 0.19,
ymax = 0.89,
] 
\addplot [blue, mark=star, mark size=2pt, mark options=solid, only marks] coordinates {
(2003,0.416666666666667)
(2004,0.25)
(2005,0.458333333333333)
(2006,0.791666666666667)
(2007,0.541666666666667)
(2008,0.458333333333333)
(2009,0.5)
(2010,0.625)
(2011,0.458333333333333)
(2012,0.25)
(2013,0.541666666666667)
(2014,0.625)
(2015,0.583333333333333)
(2016,0.416666666666667)
(2017,0.208333333333333)
(2018,0.583333333333333)
(2019,0.583333333333333)
(2020,0.625)
(2021,0.625)
(2022,0.5)
(2023,0.416666666666667)
};
\draw [red,thick,dashed] (\pgfkeysvalueof{/pgfplots/xmin},-5.0573+0.00276*\pgfkeysvalueof{/pgfplots/xmin}) -- (\pgfkeysvalueof{/pgfplots/xmax},-5.0573+0.00276*\pgfkeysvalueof{/pgfplots/xmax});
\end{axis}

\begin{axis}[
at = {(axis1.south east)},
xshift = 0.1\textwidth,
title = {Dynamic concentration $\mathit{DC}_2$},
title style = {font=\small},
xlabel = Season,
x label style = {font=\small},
x tick label style = {/pgf/number format/1000 sep={}},
ylabel = Value of competitive balance,
y label style = {font=\small},
width = 0.5\textwidth,
height = 0.4\textwidth,
ymajorgrids = true,
xmin = 2002.5,
xmax = 2023.5,
ymin = 0.19,
ymax = 0.89,
] 
\addplot [blue, mark=star, mark size=2pt, mark options=solid, only marks] coordinates {
(2003,0.4)
(2004,0.4)
(2005,0.45)
(2006,0.9)
(2007,0.55)
(2008,0.6)
(2009,0.55)
(2010,0.75)
(2011,0.4)
(2012,0.35)
(2013,0.5)
(2014,0.65)
(2015,0.7)
(2016,0.55)
(2017,0.4)
(2018,0.55)
(2019,0.6)
(2020,0.65)
(2021,0.8)
(2022,0.6)
(2023,0.45)
};
\draw [red,thick,dashed] (\pgfkeysvalueof{/pgfplots/xmin},-7.5424+0.004026*\pgfkeysvalueof{/pgfplots/xmin}) -- (\pgfkeysvalueof{/pgfplots/xmax},-7.5424+0.004026*\pgfkeysvalueof{/pgfplots/xmax});
\end{axis}
\end{tikzpicture}

\vspace{0.5cm}
\begin{tikzpicture}
\begin{axis}[
name = axis1,
title = {Dynamic concentration $\mathit{DC}_3$},
title style = {font=\small},
xlabel = Season,
x label style = {font=\small},
x tick label style = {/pgf/number format/1000 sep={}},
ylabel = Value of competitive balance,
y label style = {font=\small},
width = 0.5\textwidth,
height = 0.4\textwidth,
ymajorgrids = true,
xmin = 2002.5,
xmax = 2023.5,
ymin = 0.19,
ymax = 0.89,
] 
\addplot [blue, mark=star, mark size=2pt, mark options=solid, only marks] coordinates {
(2003,0.375)
(2004,0.416666666666667)
(2005,0.583333333333333)
(2006,0.791666666666667)
(2007,0.583333333333333)
(2008,0.625)
(2009,0.541666666666667)
(2010,0.625)
(2011,0.333333333333333)
(2012,0.291666666666667)
(2013,0.708333333333333)
(2014,0.625)
(2015,0.666666666666667)
(2016,0.458333333333333)
(2017,0.333333333333333)
(2018,0.666666666666667)
(2019,0.75)
(2020,0.666666666666667)
(2021,0.5)
(2022,0.5)
(2023,0.5)
};
\draw [red,thick,dashed] (\pgfkeysvalueof{/pgfplots/xmin},-2.9361+0.001732*\pgfkeysvalueof{/pgfplots/xmin}) -- (\pgfkeysvalueof{/pgfplots/xmax},-2.9361+0.001732*\pgfkeysvalueof{/pgfplots/xmax});
\end{axis}

\begin{axis}[
at = {(axis1.south east)},
xshift = 0.1\textwidth,
title = {Dynamic concentration $\mathit{DC}_4$},
title style = {font=\small},
xlabel = Season,
x label style = {font=\small},
x tick label style = {/pgf/number format/1000 sep={}},
ylabel = Value of competitive balance,
y label style = {font=\small},
width = 0.5\textwidth,
height = 0.4\textwidth,
ymajorgrids = true,
xmin = 2002.5,
xmax = 2023.5,
ymin = 0.19,
ymax = 0.89,
] 
\addplot [blue, mark=star, mark size=2pt, mark options=solid, only marks] coordinates {
(2003,0.35)
(2004,0.5)
(2005,0.6)
(2006,0.8)
(2007,0.55)
(2008,0.8)
(2009,0.55)
(2010,0.7)
(2011,0.25)
(2012,0.4)
(2013,0.65)
(2014,0.6)
(2015,0.6)
(2016,0.6)
(2017,0.4)
(2018,0.7)
(2019,0.85)
(2020,0.7)
(2021,0.5)
(2022,0.5)
(2023,0.45)
};
\draw [red,thick,dashed] (\pgfkeysvalueof{/pgfplots/xmin},-1.5176+0.001039*\pgfkeysvalueof{/pgfplots/xmin}) -- (\pgfkeysvalueof{/pgfplots/xmax},-1.5176+0.001039*\pgfkeysvalueof{/pgfplots/xmax});
\end{axis}
\end{tikzpicture}
\captionsetup{justification=centerfirst}
\caption{Evolution of competitive balance in the UEFA Champions League group stage \\ \vspace{0.2cm}
\footnotesize{\emph{Notes}: The seasons are indicated by their first year when the group stage is played. \\
The thick dashed red line shows the linear trend. None of them are significant even at a 25\% significance level, see Table~\ref{Table2}.}}

\label{Fig1}
\end{figure}


Figure~\ref{Fig1} shows the evolution and the linear trend of the six competitive balance indices.
The means of dynamic concentration measures are around 0.5, and even the smallest seasonal average is above 0.2. Therefore, both pot allocation and Elo ranking are better predictors of the final group ranking than a random permutation \citep{Groot2008} as expected. Group G in the 2019/20 UEFA Champions League is an interesting outlier, where the final ranking was the opposite of the ranking implied by the pot allocation.

\begin{table}[t!]
  \centering
  \caption{Linear regression models for competitive balance indicators in the \\ UEFA Champions League group stage between the 2003/04 and 2023/24 seasons}
  \label{Table2}
\rowcolors{3}{}{gray!20}
    \begin{tabularx}{\textwidth}{CcR rCC} \toprule
    Measure & Period & Intercept $c$ & Coefficient $\alpha$  & $R^2$ & $p$-value \\ \bottomrule
    $\mathit{CB}_1^A$ & 2003/04--2023/24 & $-$3.5418 & 0.001960 & 0.061 & 0.2784 \\
    $\mathit{CB}_2^A$ & 2003/04--2023/24  & $-$2.5841 & 0.001447 & 0.050 & 0.3297 \\
    $\mathit{DC}_1$ & 2003/04--2023/24  & $-$5.0573 & 0.002760 & 0.014 & 0.6047 \\
    $\mathit{DC}_2$ & 2003/04--2023/24  & $-$7.5424 & 0.004026 & 0.029 & 0.4574 \\
    $\mathit{DC}_3$ & 2003/04--2023/24  & $-$2.9361 & 0.001732 & 0.006 & 0.7460 \\
    $\mathit{DC}_4$ & 2003/04--2023/24  & $-$1.5176 & 0.001039 & 0.002 & 0.8583 \\ \bottomrule
    \end{tabularx}
\end{table}

Figure~\ref{Fig1} may suggest a small drop in competitive balance over the recent decades since the slope of the linear trend is always positive. However, none of them are significant according to Table~\ref{Table2}. This is in stark contrast with \citet{Triguero-RuizAvila-Cano2023}, where the $p$-values are below 0.02 for ex ante competitive balance and approximately 0 for dynamic concentration, respectively. Unsurprisingly, $R^2$ never exceeds 0.1, contrary to the results of \citet{RamchandaniPlumleyMondalMillarWilson2023} and \citet{Triguero-RuizAvila-Cano2023}.

\begin{table}[t!]
  \centering
  \caption{Linear regression models for ex ante competitive balance indicators in the \\ UEFA Champions League group stage between the 2003/04 and 2023/24 seasons}
  \label{Table3}
\rowcolors{3}{}{gray!20}
    \begin{tabularx}{\textwidth}{CcR rCC} \toprule
    Measure & Period & Intercept $c$ & Coefficient $\alpha$  & $R^2$ & $p$-value \\ \bottomrule
    $\mathit{CB}_1^A$ & 2003/04--2014/15 & $-$22.6219 & 0.01146 & 0.50  & 0.01034 \\
    $\mathit{CB}_1^A$ & 2014/15--2023/24 & 24.3567 & $-$0.01186 & 0.75  & 0.00124 \\
    $\mathit{CB}_2^A$ & 2003/04--2014/15 & $-$15.7854 & 0.00802 & 0.31  & 0.05861 \\
    $\mathit{CB}_2^A$ & 2014/15--2023/24 & 12.6759 & $-$0.00611 & 0.48  & 0.02656 \\ \bottomrule
    \end{tabularx}
\end{table}

Nonetheless, Table~\ref{Table3} presents that some trends can be found in ex ante competitive balance. In particular, it declined until the 2014/15 season and improved between the 2014/15 and 2023/24 seasons. The pattern is less robust but still exists if only the three strongest teams are considered in all groups as index $CB_2^A$ does.

\section{Discussion} \label{Sec6}

How can these divergent conclusions be explained?
Although both \citet{RamchandaniPlumleyMondalMillarWilson2023} (28 seasons from 1992/93 to 2019/20) and \citet{Triguero-RuizAvila-Cano2023} (19 seasons from 1999/2000 to 2017/2018) study different samples, this cannot be the main reason. According to \citet{RamchandaniPlumleyMondalMillarWilson2023}, the reduction in ex post competitive balance is even more evident between the 2002/03 and 2019/20 seasons, where they report $R^2 = 0.49$ for the linear regression; unfortunately, the $p$-value is not reported. Analogously, the five-year averages of the DCB index continuously increase---the competitive balance worsens---in both the ex ante and ex post settings \citep{Triguero-RuizAvila-Cano2023}.
Nonetheless, Table~\ref{Table3} suggests that the improvement in the recent seasons not considered in previous studies is partially responsible for the insignificant trend found here.

Our ex ante competitive balance indicators are based on the more accurate Elo rating. \citet{Triguero-RuizAvila-Cano2023} use the UEFA club coefficients, which have likely become more concentrated among the leading clubs without a parallel dominance in the Elo ratings. A potential reason might be that the best clubs in the top leagues---except England, where the set of competitors is wider---can somewhat relax in their domestic leagues now since the number of guaranteed slots in the UEFA Champions League group stage has increased from two to four for the four highest-ranked leagues. Therefore, they are able to increasingly focus on international competitions.

Regarding dynamic concentration, we compare the final group ranking to a benchmark provided by either the pot allocation or the Elo ranking. Since $\mathit{HHI}$ does not require such a benchmark, a worsening competitive balance may be covered if our benchmark becomes a more accurate predictor over time. However, this is improbable. Even though the seeding system has been reformed as discussed in Section~\ref{Sec3}, it has rather increased uncertainty \citep{CoronaForrestTenaHorrilloWiper2019, DagaevRudyak2019}. Similarly, the differences in the Elo ratings have not increased (Figure~\ref{Fig1}), thus, the quality of Elo ranking seems to be unchanged. On the other hand, the clubs may focus in the group stage more strongly on their number of points than previously, possibly due to the increasing financial incentives provided by the UEFA. This could lead to a higher $\mathit{HHI}$ and a lower probability of losing stakeless games. Then the earlier studies have mixed a favourable trend (teams do not lose their stakeless matches) with declining ex post competitive balance. The hypothesis may be tested by an empirical investigation of stakeless games, which can be identified, for example, by the approach of \citet{CsatoMolontayPinter2024}.
Finally, it should be noted that the sample size is rather small. However, the novel format of the UEFA Champions League from the 2024/25 season does not allow for more observations.

\section{Conclusions} \label{Sec7}

UEFA has chosen a risky strategy by fundamentally reforming the well-established group stage of its major tournament, the UEFA Champions League, from the 2024/25 season. According to the existing literature based on straightforward $\mathit{HHI}$-based measures, UEFA should have acted effectively because of the notable drop in both the ex ante and ex post competitive balance of the group stage over the last two decades.

Our paper has proposed six alternative but reasonable indices to investigate the evolution of competitive balance in the UEFA Champions League group stage. Crucially, no evidence is found for any decline between the 2003/04 and 2023/24 seasons. Therefore, we urge researchers and analysts to consider a broader set of competitive balance indicators in the future before proposing powerful policy interventions.

\section*{Acknowledgements}
\addcontentsline{toc}{section}{Acknowledgements}
\noindent
This paper could not have been written without \emph{R\'eka Boros} and \emph{Adrienn Czak\'o}, who prepared the data. \\
We are indebted to \emph{Andr\'as Gyimesi} for useful suggestions. \\
One anonymous reviewer provided valuable remarks on earlier drafts. \\
The research was supported by the National Research, Development and Innovation Office under Grants FK 145838 and PD 153835, and by the J\'anos Bolyai Research Scholarship of the Hungarian Academy of Sciences.

\bibliographystyle{apalike}
\bibliography{All_references}

\begin{thebibliography}{}

\bibitem[Ausloos, 2024a]{Ausloos2024a}
Ausloos, M. (2024a).
\newblock Hierarchy selection: New team ranking indicators for cyclist
  multi-stage races.
\newblock {\em European Journal of Operational Research}, 314(2):807--816.

\bibitem[Ausloos, 2024b]{Ausloos2024b}
Ausloos, M. (2024b).
\newblock Should one (be allowed to) replace the {C}ippolini's?
\newblock {\em Annals of Operations Research}, in press.
\newblock {DOI}:
  \href{https://doi.org/10.1007/s10479-024-06206-y}{10.1007/s10479-024-06206-y}.

\bibitem[Avila-Cano et~al., 2023]{Avila-CanoOwenTriguero-Ruiz2023}
Avila-Cano, A., Owen, P.~D., and Triguero-Ruiz, F. (2023).
\newblock Measuring competitive balance in sports leagues that award bonus
  points, with an application to rugby union.
\newblock {\em European Journal of Operational Research}, 309(2):939--952.

\bibitem[Avila-Cano et~al., 2021]{Avila-CanoRuiz-SepulvedaTriguero-Ruiz2021}
Avila-Cano, A., Ruiz-Sepulveda, A., and Triguero-Ruiz, F. (2021).
\newblock Identifying the maximum concentration of results in bilateral sports
  competitions.
\newblock {\em Mathematics}, 9(11):1293.

\bibitem[Basini et~al., 2023]{BasiniTsouliNtzoufrasFriel2023}
Basini, F., Tsouli, V., Ntzoufras, I., and Friel, N. (2023).
\newblock Assessing competitive balance in the {E}nglish {P}remier {L}eague for
  over forty seasons using a stochastic block model.
\newblock {\em Journal of the Royal Statistical Society Series A: Statistics in
  Society}, 186(3):530--556.

\bibitem[Bell, 2018]{Bell2018}
Bell, A. (2018).
\newblock Real a shambles in {CSKA} loss.
\newblock 12 December.
  \url{https://www.marca.com/en/football/real-madrid/2018/12/12/5c116ac522601d22118b45c5.html}.

\bibitem[Bosker and G{\"u}rtler, 2024]{BoskerGurtler2024}
Bosker, J. and G{\"u}rtler, M. (2024).
\newblock The impact of cultural differences on the success of elite labor
  migration---{E}vidence from professional soccer.
\newblock {\em Annals of Operations Research}, 341(2-3):781--824.

\bibitem[Buraimo et~al., 2022]{BuraimoForrestMcHaleTena2022}
Buraimo, B., Forrest, D., McHale, I.~G., and Tena, J.~D. (2022).
\newblock Armchair fans: Modelling audience size for televised football
  matches.
\newblock {\em European Journal of Operational Research}, 298(2):644--655.

\bibitem[Chater et~al., 2021]{ChaterArrondelGayantLaslier2021}
Chater, M., Arrondel, L., Gayant, J.-P., and Laslier, J.-F. (2021).
\newblock Fixing match-fixing: Optimal schedules to promote competitiveness.
\newblock {\em European Journal of Operational Research}, 294(2):673--683.

\bibitem[Corona et~al., 2019]{CoronaForrestTenaHorrilloWiper2019}
Corona, F., Forrest, D., Tena~Horrillo, J.~{\relax de~D}., and Wiper, M.
  (2019).
\newblock Bayesian forecasting of {UEFA} {C}hampions {L}eague under alternative
  seeding regimes.
\newblock {\em International Journal of Forecasting}, 35(2):722--732.

\bibitem[Csat{\'o}, 2020]{Csato2020a}
Csat{\'o}, L. (2020).
\newblock The {UEFA} {C}hampions {L}eague seeding is not strategy-proof since
  the 2015/16 season.
\newblock {\em Annals of Operations Research}, 292(1):161--169.

\bibitem[Csat{\'o}, 2021]{Csato2021a}
Csat{\'o}, L. (2021).
\newblock {\em Tournament Design: How Operations Research Can Improve Sports
  Rules}.
\newblock Palgrave Pivots in Sports Economics. Palgrave Macmillan, Cham,
  Switzerland.

\bibitem[Csat{\'o}, 2022]{Csato2022b}
Csat{\'o}, L. (2022).
\newblock {UEFA} against the champions? {A}n evaluation of the recent reform of
  the {C}hampions {L}eague qualification.
\newblock {\em Journal of Sports Economics}, 23(8):991--1016.

\bibitem[Csat{\'o}, 2024]{Csato2024c}
Csat{\'o}, L. (2024).
\newblock Club coefficients in the {UEFA} {C}hampions {L}eague: Time for shift
  to an {E}lo-based formula.
\newblock {\em International Journal of Performance Analysis in Sport},
  24(2):119--134.

\bibitem[Csat{\'o} et~al.,
  2025]{CsatoDevriesereGoossensGyimesiLambersSpieksma2025}
Csat{\'o}, L., Devriesere, K., Goossens, D., Gyimesi, A., Lambers, R., and
  Spieksma, F. (2025).
\newblock Ranking matters: Does the new format select the best teams for the
  knockout phase in the {UEFA} {C}hampions {L}eague?
\newblock Manuscript. {DOI}:
  \href{https://doi.org/10.48550/arXiv.2503.13569}{10.48550/arXiv.2503.13569}.

\bibitem[Csat{\'o} and Ilyin, 2025]{CsatoIlyin2025}
Csat{\'o}, L. and Ilyin, S. (2025).
\newblock Misaligned incentives in sports: A mathematical analysis of the
  post-2024 {UEFA} {C}hampions {L}eague qualification.
\newblock {\em IMA Journal of Management Mathematics}, dpaf016.

\bibitem[Csat{\'o} et~al., 2024]{CsatoMolontayPinter2024}
Csat{\'o}, L., Molontay, R., and Pint{\'e}r, J. (2024).
\newblock Tournament schedules and incentives in a double round-robin
  tournament with four teams.
\newblock {\em International Transactions in Operational Research},
  31(3):1486--1514.

\bibitem[Csat{\'o} and Petr{\'o}czy, 2024]{CsatoPetroczy2024}
Csat{\'o}, L. and Petr{\'o}czy, D.~G. (2024).
\newblock Bibliometric indices as a measure of performance and competitive
  balance in the knockout stage of the {UEFA} {C}hampions {L}eague.
\newblock {\em Central European Journal of Operations Research},
  32(4):8961--988.

\bibitem[Dagaev and Rudyak, 2019]{DagaevRudyak2019}
Dagaev, D. and Rudyak, V. (2019).
\newblock Seeding the {UEFA} {C}hampions {L}eague participants: evaluation of
  the reforms.
\newblock {\em Journal of Quantitative Analysis in Sports}, 15(2):129--140.

\bibitem[Depken, 1999]{Depken1999}
Depken, C.~A. (1999).
\newblock Free-agency and the competitiveness of {M}ajor {L}eague {B}aseball.
\newblock {\em Review of Industrial Organization}, 14(3):205--217.

\bibitem[Devriesere et~al., 2025]{DevriesereCsatoGoossens2025}
Devriesere, K., Csat{\'o}, L., and Goossens, D. (2025).
\newblock Tournament design: A review from an operational research perspective.
\newblock {\em European Journal of Operational Research}, 324(1):1--21.

\bibitem[Doria and Nalebuff, 2021]{DoriaNalebuff2021}
Doria, M. and Nalebuff, B. (2021).
\newblock Measuring competitive balance in sports.
\newblock {\em Journal of Quantitative Analysis in Sports}, 17(1):29--46.

\bibitem[Dubois, 2022]{Dubois2022}
Dubois, M. (2022).
\newblock Dominance criteria on grids for measuring competitive balance in
  sports leagues.
\newblock {\em Mathematical Social Sciences}, 115:1--10.

\bibitem[El-Hodiri and Quirk, 1971]{El-HodiriQuirk1971}
El-Hodiri, M. and Quirk, J. (1971).
\newblock An economic model of a professional sports league.
\newblock {\em Journal of Political Economy}, 79(6):1302--1319.

\bibitem[Ely et~al., 2015]{ElyFrankelKamenica2015}
Ely, J., Frankel, A., and Kamenica, E. (2015).
\newblock Suspense and surprise.
\newblock {\em Journal of Political Economy}, 123(1):215--260.

\bibitem[{Football Club Elo Ratings}, 2024]{FootballClubEloRatings}
{Football Club Elo Ratings} (2024).
\newblock The system or how this works.
\newblock \url{http://clubelo.com/System}.

\bibitem[Fort and Maxcy, 2003]{FortMaxcy2003}
Fort, R. and Maxcy, J. (2003).
\newblock Competitive balance in sports leagues: An introduction.
\newblock {\em Journal of Sports Economics}, 4(2):154--160.

\bibitem[Gerrard and Kringstad, 2022]{GerrardKringstad2022}
Gerrard, B. and Kringstad, M. (2022).
\newblock The multi-dimensionality of competitive balance: Evidence from
  {E}uropean football.
\newblock {\em Sport, Business and Management: An International Journal},
  12(4):382--402.

\bibitem[Gerrard and Kringstad, 2023]{GerrardKringstad2023}
Gerrard, B. and Kringstad, M. (2023).
\newblock Dispersion and persistence in the competitive balance of {N}orth
  {A}merican major leagues 1960--2019.
\newblock {\em Sport, Business and Management: An International Journal},
  13(5):640--662.

\bibitem[Groot, 2008]{Groot2008}
Groot, L. (2008).
\newblock {\em Economics, Uncertainty and European Football: Trends in
  Competitive Balance}.
\newblock New Horizons in the Economics of Sport. Edward Elgar Publishing,
  Cheltenham, United Kigndom.

\bibitem[Guyon, 2015]{Guyon2015a}
Guyon, J. (2015).
\newblock Rethinking the {FIFA} {W}orld {C}up\textsuperscript{{TM}} final draw.
\newblock {\em Journal of Quantitative Analysis in Sports}, 11(3):169--182.

\bibitem[Guyon, 2020]{Guyon2020a}
Guyon, J. (2020).
\newblock Risk of collusion: {W}ill groups of 3 ruin the {FIFA} {W}orld {C}up?
\newblock {\em Journal of Sports Analytics}, 6(4):259--279.

\bibitem[Guyon, 2022]{Guyon2022a}
Guyon, J. (2022).
\newblock ``{C}hoose your opponent'': A new knockout design for hybrid
  tournaments.
\newblock {\em Journal of Sports Analytics}, 8(1):9--29.

\bibitem[Gyimesi, 2024]{Gyimesi2024}
Gyimesi, A. (2024).
\newblock Competitive balance in the post-2024 {C}hampions {L}eague and the
  {E}uropean {S}uper {L}eague: {A} simulation study.
\newblock {\em Journal of Sports Economics}, 25(6):707--734.

\bibitem[Herfindahl, 1950]{Herfindahl1950}
Herfindahl, O.~C. (1950).
\newblock {\em Concentration in the Steel Industry}.
\newblock PhD thesis, Columbia University, New York.

\bibitem[Hirschman, 1945]{Hirschman1945}
Hirschman, A.~O. (1945).
\newblock {\em National {P}ower and the {S}tructure of {F}oreign {T}rade}.
\newblock University of California Press, Berkeley and Los Angeles, California,
  USA.

\bibitem[Hirschman, 1964]{Hirschman1964}
Hirschman, A.~O. (1964).
\newblock The paternity of an index.
\newblock {\em American Economic Review}, 54(5):761--762.

\bibitem[Hood and Jewell, 2022]{HoodJewell2022}
Hood, M. and Jewell, R.~T. (2022).
\newblock What does it mean to be ``competitive''? {U}sing simulation to reveal
  competitive balance in {E}nglish {A}ssociation {F}ootball.
\newblock {\em International Journal of Sport Finance}, 17(3):125--139.

\bibitem[Humphreys, 2002]{Humphreys2002}
Humphreys, B.~R. (2002).
\newblock Alternative measures of competitive balance in sports leagues.
\newblock {\em Journal of Sports Economics}, 3(2):133--148.

\bibitem[Kendall and Lenten, 2017]{KendallLenten2017}
Kendall, G. and Lenten, L.~J.~A. (2017).
\newblock When sports rules go awry.
\newblock {\em European Journal of Operational Research}, 257(2):377--394.

\bibitem[Kendall, 1938]{Kendall1938}
Kendall, M.~G. (1938).
\newblock A new measure of rank correlation.
\newblock {\em Biometrika}, 30(1/2):81--93.

\bibitem[Koning, 2009]{Koning2009}
Koning, R.~H. (2009).
\newblock Sport and measurement of competition.
\newblock {\em De Economist}, 157(2):229--249.

\bibitem[Kringstad and Gerrard, 2007]{KringstadGerrard2007}
Kringstad, M. and Gerrard, B. (2007).
\newblock Beyond competitive balance.
\newblock In Parent, M.~M. and Slack, T., editors, {\em International
  Perspectives on the Management of Sport}, pages 149--172. Elsevier, London,
  United Kingdom.

\bibitem[Laliena and L{\'o}pez, 2019]{LalienaLopez2019}
Laliena, P. and L{\'o}pez, F.~J. (2019).
\newblock Fair draws for group rounds in sport tournaments.
\newblock {\em International Transactions in Operational Research},
  26(2):439--457.

\bibitem[Manasis et~al., 2011]{ManasisAvgerinouNtzoufras2011}
Manasis, V., Avgerinou, V., Ntzoufras, I., and Reade, J.~J. (2011).
\newblock Measurement of competitive balance in professional team sports using
  the {N}ormalized {C}oncentration {R}atio.
\newblock {\em Economics Bulletin}, 31(3):2529--2540.

\bibitem[Manasis et~al., 2013]{ManasisAvgerinouNtzoufrasReade2013}
Manasis, V., Avgerinou, V., Ntzoufras, I., and Reade, J.~J. (2013).
\newblock Quantification of competitive balance in {E}uropean football:
  development of specially designed indices.
\newblock {\em IMA Journal of Management Mathematics}, 24(3):363--375.

\bibitem[Manasis and Ntzoufras, 2014]{ManasisNtzoufras2014}
Manasis, V. and Ntzoufras, I. (2014).
\newblock Between-seasons competitive balance in {E}uropean football: review of
  existing and development of specially designed indices.
\newblock {\em Journal of Quantitative Analysis in Sports}, 10(2):139--152.

\bibitem[Manasis et~al., 2022]{ManasisNtzoufrasReade2022}
Manasis, V., Ntzoufras, I., and Reade, J.~J. (2022).
\newblock Competitive balance measures and the uncertainty of outcome
  hypothesis in {E}uropean football.
\newblock {\em IMA Journal of Management Mathematics}, 33(1):19--52.

\bibitem[Noll, 1988]{Noll1988}
Noll, R.~G. (1988).
\newblock Professional basketball.
\newblock Studies in Industrial Economics Paper No.~144., Stanford University
  Press, Stanford, California, USA.

\bibitem[Owen and Owen, 2022]{OwenOwen2022}
Owen, P.~D. and Owen, C.~A. (2022).
\newblock Simulation evidence on {H}erfindahl-{H}irschman measures of
  competitive balance in professional sports leagues.
\newblock {\em Journal of the Operational Research Society}, 73(2):285--300.

\bibitem[Owen et~al., 2007]{OwenRyanWeatherston2007}
Owen, P.~D., Ryan, M., and Weatherston, C.~R. (2007).
\newblock Measuring competitive balance in professional team sports using the
  {H}erfindahl-{H}irschman index.
\newblock {\em Review of Industrial Organization}, 31(4):289--302.

\bibitem[Pawlowski and Nalbantis, 2019]{PawlowskiNalbantis2019}
Pawlowski, T. and Nalbantis, G. (2019).
\newblock Competitive balance: Measurement and relevance.
\newblock In Downward, P., Frick, B., Humphreys, B.~R., Pawlowski, T., Ruseski,
  J.~E., and Soebbing, B.~P., editors, {\em The Sage Handbook of Sports
  Economics}, pages 154--171. Sage Publications, London, United Kingdom.

\bibitem[Ramchandani et~al., 2023]{RamchandaniPlumleyMondalMillarWilson2023}
Ramchandani, G., Plumley, D., Mondal, S., Millar, R., and Wilson, R. (2023).
\newblock `{Y}ou can look, but don't touch': competitive balance and dominance
  in the {UEFA} {C}hampions {L}eague.
\newblock {\em Soccer \& Society}, 24(4):479--491.

\bibitem[Richardson et~al., 2023]{RichardsonNalbantisPawlowski2023}
Richardson, T., Nalbantis, G., and Pawlowski, T. (2023).
\newblock Emotional cues and the demand for televised sports: {E}vidence from
  the {UEFA} {C}hampions {L}eague.
\newblock {\em Journal of Sports Economics}, 24(8):993--1025.

\bibitem[Rottenberg, 1956]{Rottenberg1956}
Rottenberg, S. (1956).
\newblock The baseball players' labor market.
\newblock {\em Journal of Political Economy}, 64(3):242--258.

\bibitem[Sanderson, 2002]{Sanderson2002}
Sanderson, A.~R. (2002).
\newblock The many dimensions of competitive balance.
\newblock {\em Journal of Sports Economics}, 3(2):204--228.

\bibitem[Scelles et~al., 2022]{ScellesFrancoisDermit-Richard2022}
Scelles, N., Fran{\c{c}}ois, A., and Dermit-Richard, N. (2022).
\newblock Determinants of competitive balance across countries: Insights from
  {E}uropean men's football first tiers, 2006--2018.
\newblock {\em Managing Sport and Leisure}, 27(3):267--284.

\bibitem[Scully, 1989]{Scully1989}
Scully, G.~W. (1989).
\newblock {\em The Business of Major League Baseball}.
\newblock University of Chicago Press, Chicago, Illinois, USA.

\bibitem[Triguero~Ruiz and Avila-Cano, 2019]{Triguero-RuizAvila-Cano2019}
Triguero~Ruiz, F. and Avila-Cano, A. (2019).
\newblock The distance to competitive balance: A cardinal measure.
\newblock {\em Applied Economics}, 51(7):698--710.

\bibitem[Triguero-Ruiz and Avila-Cano, 2023]{Triguero-RuizAvila-Cano2023}
Triguero-Ruiz, F. and Avila-Cano, A. (2023).
\newblock On competitive balance in the group stage of the {UEFA} {C}hampions
  {L}eague.
\newblock {\em Scottish Journal of Political Economy}, 70(3):231--248.

\bibitem[Triguero-Ruiz et~al., 2023]{Triguero-RuizOwenAvila-Cano2023}
Triguero-Ruiz, F., Owen, P.~D., and Avila-Cano, A. (2023).
\newblock The minimum concentration of points in sports leagues without ties.
\newblock {\em Sports Economics Review}, 4:100019.

\bibitem[UEFA, 2022]{UEFA2022a}
UEFA (2022).
\newblock {UEFA} approves final format and access list for its club
  competitions as of the 2024/25 season.
\newblock 10 May.
  \url{https://www.uefa.com/returntoplay/news/0275-151c779310c3-b92bbf0d24f9-1000--uefa-approves-final-format-and-access-list-for-its-club-competi/}.

\bibitem[van Ours, 2024]{vanOurs2024}
van Ours, J.~C. (2024).
\newblock Nontransitive patterns in long-term football rivalries.
\newblock {\em Journal of Sports Economics}, 25(7):802--826.

\bibitem[Yildirim and Bilman, 2025a]{YildirimBilman2025a}
Yildirim, M. and Bilman, M.~E. (2025a).
\newblock New rules, new game? {T}he effects of the away goals rule removal and
  video assistant referee adoption on game dynamics in {UEFA} {C}hampions
  {L}eague ties.
\newblock {\em Journal of Policy Modeling}, 47(1):78--96.

\bibitem[Yildirim and Bilman, 2025b]{YildirimBilman2025b}
Yildirim, M. and Bilman, M.~E. (2025b).
\newblock Second-leg home advantage no more? {T}he impact of {V}ideo
  {A}ssistant {R}eferee and no away goals rule in elite soccer.
\newblock {\em Journal of Policy Modeling}, in press.
\newblock {DOI}:
  \href{https://doi.org/10.1016/j.jpolmod.2025.06.003}{10.1016/j.jpolmod.2025.06.003}.

\bibitem[Zimbalist, 2002]{Zimbalist2002}
Zimbalist, A.~S. (2002).
\newblock Competitive balance in sports leagues: An introduction.
\newblock {\em Journal of Sports Economics}, 3(2):111--121.

\end{thebibliography}

\end{document}